\documentclass [manuscript] {acmart}

\AtBeginDocument{%
  \providecommand\BibTeX{{%
    \normalfont B\kern-0.5em{\scshape i\kern-0.25em b}\kern-0.8em\TeX}}}

\setcopyright{acmcopyright}
\copyrightyear{2018}
\acmYear{2018}
\acmDOI{XXXXXXX.XXXXXXX}

\acmJournal{JACM}
\acmVolume{37}
\acmNumber{4}
\acmArticle{111}
\acmMonth{8}

\acmPrice{15.00}
\acmISBN{978-1-4503-XXXX-X/18/06}

\begin{document}

\title{In the user's eyes we find trust: Using gaze data as a predictor or trust in an artifical intelligence}

\author{Martin Dechant}
\email{martin.dechant@zeiss.com}
\orcid{0000-0001-9073-8727}
\affiliation{%
  \institution{Carl Zeiss Vision International GmbH}
  \streetaddress{Turnstrasse 27}
  \city{Aalen}
  \country{Germany}
  \postcode{73430}
}

\author{Olga Lukashova-Sanz}
\email{olga.lukashova@zeiss.com}
\orcid{0000-0002-4053-3894}
\affiliation{%
  \institution{Carl Zeiss Vision International GmbH}
  \streetaddress{Turnstrasse 27}
  \city{Aalen}
  \country{Germany}
  \postcode{73430}
}

\author{Siegfried Wahl}
\email{siegfried.wahl@zeiss.com}
\orcid{0000-0003-3437-6711}
\affiliation{%
  \institution{Carl Zeiss Vision International GmbH}
  \streetaddress{Turnstrasse 27}
  \city{Aalen}
  \country{Germany}
  \postcode{73430}
}

\renewcommand{\shortauthors}{Dechant and Wahl, et al.}

\begin{abstract}
Trust is essential for our interactions with others but also with artificial intelligence (AI) based systems. To understand whether a user trusts an AI, researchers need reliable measurement tools. However, currently discussed markers mostly rely on expensive and invasive sensors, like electroencephalograms, which may cause discomfort. The analysis of gaze data has been suggested as a convenient tool for trust assessment. However, the relationship between trust and several aspects of the gaze behaviour is not yet fully understood. To provide more insights into this relationship, we propose a exploration study in virtual reality where participants have to perform a sorting task together with a simulated AI in a simulated robotic arm embedded in a gaming. We discuss the potential benefits of this approach and outline our study design in this submission.

\end{abstract}



\maketitle

\section{Introduction}
Just as trust is an important component of interactions between humans, trust plays a pivotal role when interacting with a device or even algorithm  \cite{Arrow1974TheOrganization,Liu2010Human-MachineInteraction}. AI-based systems have been improving the quality of many lives in various situations like online shopping \cite{Bawack2022ArtificialReview}, media consumption \cite{Lachman2021ApplicationsEntertainment} and entertainment industry \cite{Chan-Olmsted2019AIndustry}, but also in the realm of health care and well-being \cite{Yu2018ArtificialHealthcare}. As prior work shows, trust is a major antecedent of technology acceptance and usage \cite{Choung2022TrustTechnologiesb}. Without it, users may experience unpleasant emotions, such as anxiety\cite{Johnson2017AIAnxiety}, while interacting with a system which they don't trust or even stop using it due to the lack of trust in it. Like other technologies, AI faces the ongoing challenge of earning the user's trust \cite{Glikson2020HumanResearch,Bockle2021CanInterfaces}: On the other hand, while AI based systems can improve the quality of life for many users, there is an ongoing public debate about trust in the systems as well as the procedures linked to AI. 

With the increasing interest in enhancing our daily lives with artificial intelligent based approaches and the underlying question about trust in these systems, there is an increased demand for understanding and measuring user's trust in artificial intelligent systems as well as the factors that may affect the formation of trust \cite{Brzowski2019TrustReview}. In recent years the researchers began to address this demand by several approaches for trust measurements in several ways: First, providing a suitable trust definition and the declaration which aspects of the human-AI interaction are important for the formation of trust. 
Second, creation and validation of standardized questionnaires for trust for easy comparison between different projects and underlying ideas. Third: development of additional measurements besides questionnaire and rating based approaches for assessing the user's trust in the moment of the interaction. 

One particular aspect in these efforts is the development of (digital) biomarker \cite{Strimbu2010,Dorsey2017} which may help to better understand \textit{when} the user is experiencing trust in a system. Behaviour-based approaches may be deployed without distracting the user from the current task \cite{Ezer2019TrustTeams},and reduce the risk related to self-reported information, such as the white coat effect \cite{Shehab2011CognitiveHypertension}. Similar to other mental state assessments, like the assessment of anxiety disorders \cite{Antony2005}, the combination of measurements about the user's self-perspective as well as the analysis of their behaviour while interacting with the system my deliver more robust insights about trust during a human-AI-interaction.

Therefore researchers analyzed trust-related behaviour as a bridge towards measuring trust. In this work certain behaviours \cite{Vereschak2021HowMethodologies}, such as extended decision time  \cite{Yuksel2017BrainsInteractions,Feng2019WhatPlay},the user's compliance with the AI's suggestions, and switch ratio between accepting and rejecting the AI's suggestion (see \cite{Vereschak2021HowMethodologies} for details) where identified as potential trust-related behaviour. Furthermore, physiological measurements of the user's skin conductance, heartbeat, and brain activities, and the user's mouse movement were introduced as additional measurements for trust \cite{Freeman2018DoingHand}. 

Prior work successfully used gaze data to gain more insights about bottom-up and top-down cognitive processes as well as the user's engagement with a task \cite{Eckstein2017BeyondDevelopment}. In the context of trust, prior work in the robot-human interaction as well as in the driver assistance show that gaze may be a handy tool for measuring trust. However, most of these projects focused on the user's attention, expressed through the fixation times during an experiment. Other aspects of the user's gaze data, like saccades, may be a rich source for analyzing the user's trust in an artificial intelligence.  

Therefore, we argue that gaze data may be a valuable additional tool to enhance existing techniques: Recording the user's gaze needs a less complex setup in comparison to other sensors (e.g. EEG, Skin conductance) \cite{Vereschak2021HowMethodologies}, has proven to be a reliable data source in various other research projects. Furthermore early work shows that the user's focus plays an essential role in determining whether the user trusts the system or not. Also, eye tracking data has been used to measure the cognitive workload in adjunct fields which may be a useful digital biomarker \cite{Barbato2020TheOrienting,Zhang2020HumanReview}. To better understand the relationship between the user's gaze and the experienced trust in a system, we propose a study where a user has to sort together with an automated robot arm together embedded in a gaming task. We present related work, propose our study design and discuss potential challenges and benefits of gaze-based trust assessment.

\section{Related Work}
The attitude characteristic of trust makes a reliable assessment of the user's trust in a system quite challenging \cite{Papenmeier2022ItsAI,Yi2020IdentificationEmotions}. This means that an individual may experience trust, but not reveal the experience of it through behaviour. To overcome this challenge, researchers apply different approaches and technologies ranging from the use of standardized questionnaires to the measurement of behaviour linked to potential trustful interaction \cite{Vereschak2021HowMethodologies}. However, there is still an ongoing debate about whether such approaches are valid \cite{Papenmeier2022ItsAI,Vereschak2021HowMethodologies}. Different contexts may require different aspects of trust to be measured. As previously discussed: trust can either focus on a device in a specific moment or the general trust in a type of technology \cite{Wrightsman_1991,Merritt2013ISystem}.

\subsection{Questionnaire-Based and Subjective Assessment of Trust}
Standardized questionnaires and ratings are one commonly used way to assess the subjective view of the users and allow them to rate whether they experience trust. Most of these questionnaires depend on the context and the underlying research question. For example: \citet{Merritt2013ISystem} focuses on the overall trust in technology, while other questionnaires, like that of \citet{Wrightsman_1991}, focus on whether an individual generally trusts other people \cite{Wrightsman_1991}. Another way to assess trust is a simple rating measurement, where users simply indicate how much they trusted a system \cite{Yin2019UnderstandingModels}. However, some researchers argue that these questionnaires and rating-based measurements of trust are not able to capture the complexity of trust \cite{Loo2002AScales}. Other researchers argue that trust should not only be measured on its own and recommend combining additional measurements to distinguish between adjunct phenomena \cite{Vereschak2021HowMethodologies}.  

\subsection{Behavioural Markers and Physiological Data for Assessing Trust}
Besides the use of questionnaires, researchers have started to harness additional data sources to find ways to assess trust in different contexts, including the human-AI interaction \cite{HarrisonMcKnight2001TrustTime}. However, these may be among the most affected by the characteristics of trust, meaning that trust may not be directly expressed through measurable behaviour or signals \cite{Castelfranchi2010_Socio}. In comparison: emotional states such as anxiety may cause distinct physical reactions \cite{Spence2016} such as trembling \cite{Connor2000PsychometricScale} or increased pulse \cite{Cheng2022HeartMetaanalysis}. Therefore, prior work suggests \cite{HarrisonMcKnight2001TrustTime,Castelfranchi2010_Socio,Vereschak2021HowMethodologies} describing these measurements as trust-related behavioural measurements to account for the fact that these measurements may not fully represent trust. 

\subsection{The Usage of Gaze Data for Assessing Cognitive Characteristics}
The human's gaze contains many information about internal cognitive states and brain activities while being non-invasive \cite{Zhang2020HumanReview}. In comparison to other technologies, such as EEG and fMRI, gaze tracking technology is cheaper, portable and therefore allows a more diverse research approach. Furthermore, the rapid calibration and late improvements in gaze tracking approaches allow a faster experiment design due to a faster calibration of the sensors \cite{Eckstein2017BeyondDevelopment}.  
Many researchers use gaze data as a way to measure the user's attention during an experiment to ensure they stay focused on the task \cite{Theeuwes2009InteractionsMovements,Yarbus1967EyeObjects}. Furthermore, multiple gaze metrics have been used to study cognition of users which are derived from eye position data: Fixation, the time how long users are fixating a certain point with their gaze \cite{Eckstein2017BeyondDevelopment}, has been applied to gain insights about the formation of memories \cite{Theeuwes2009InteractionsMovements}, the mental computations \cite{Green2007EyeAddition}, reading \cite{Rayner1998EyeResearch.}, learning as well as problem solving \cite{Lai2013A2012,Grant2003EyeThought}. Furthermore, the analysis of saccades, the rapid shift between fixations, has been applied to gain insights about the user's focus and distractions \cite{Eckstein2017BeyondDevelopment}. Especially the accuracy and latency has revealed insights about cognitive control capacities \cite{Luna2011DevelopmentMovements}. Furthermore, by analyzing the scan path, researchers were able to distinct novices and experts in various fields \cite{Koffskey2014UsingOperators,Nazareth2020TheTechnology} which shows the many ways eye tracking may be useful for gaining insights about the user's cognitive state. 
The breadth of cognitive insights hidden in the user's gaze behaviour motivated to expand currently discussed eye-tracking methodology. For example: Recent discussions around the analysis of the pupil dilation \cite{Barreto2008} as well as the blink rate have been proposed as ways to analyze the cognitive work load as well as the fatigue of the users \cite{Paprocki2017WhatPerformance}.
Also within the context of trust measurement through gaze, first work in the context of trust in automated driving show that the user's attention shifts and the underlying gaze behaviour may reveal whether the user trust a system or not \cite{Du2020PsychophysiologicalDriving}. By focusing more often on information about the automation users may reveal that they do not trust the system and prepare themselves to intervene. 

Overall, first evidence suggests that there is a relationship between the user's attention and especially the attention shifts and the experience of trust \cite{Bayliss2006PredictiveYou}. However, several factors, such as demographics as well as personal experience may affect these attention shifts. Furthermore, other aspects, especially the pupil and other aspects of the eye movement, like a closer analysis of saccade's characteristics, may reveal additional information about whether the user is experiencing trust in an artificial intelligence embedded in a system. 

\section{Planned Experiment}
Therefore, we plan to conduct a virtual reality study to investigate the relationship between the subjective experience of trust and the user's gaze behaviour during a collaborative gaming task. Based on prior work we included two conditions of the experiment based on the accuracy of the AI: In one condition the AI will make more errors and in the second condition the AI will make almost no errors. This will step may help us to better understand the role of the AI's performance on the experience of trust as well as how it affects the user's gaze.

\subsection{Gaming Task}
\begin{figure}[htp]
\includegraphics[width=0.9\textwidth]{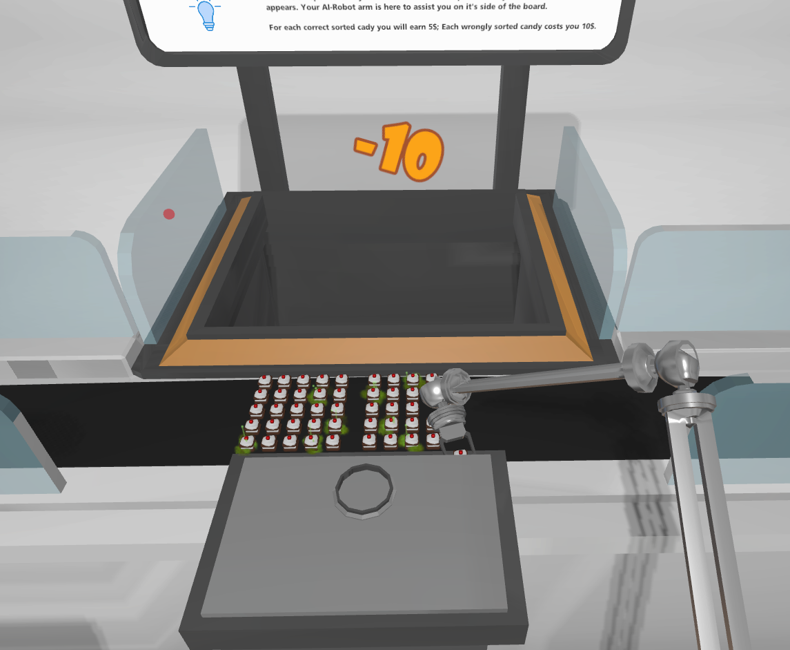}
\centering
\caption{Screenshot of the prototype: A robot arm, controlled by the AI and the player have to sort spoiled candy in front of them. The right side is only interactable for the AI, the left side for the user. Both, the AI and the player have to place their spoiled candy in the bin behind the conveyor belt. For \textbf{each} correctly placed candy, the player earns 5 points and for each wrongly placed the player looses 10 points.}
\Description{&&&}
\label{fig:Prototype}
\end{figure}

First we introduce the game's story and the control scheme: As a new intern at a large candy factory participants are instructed to work as quality assurance engineer at a conveyor belt. There participants have to sort out spoiled candy from a box. This candy box is divided into two halves: One side is interactable for the player, the other side is managed by a robotic arm and therefore not directly interactable for the user. Both sides are highlighted through different colours. Once the introduction and the eye tracker calibration is finished, players are instructed to press a button in front of them in the virtual environment for 3 seconds. After that the candy box appears in front of the player. Each side has 4 pieces of spoiled candy which have to be removed. The robot arm will automatically start to grab 4 pieces from it's side and place them in the trash bin behind the conveyor belt. Similar to that, players are also instructed to grab spoiled candy and place them in the same trash bin as the robotic arm. For each correct grabbed and removed candy, players score 5 points. Each time either the robotic arm or the player places a wrong candy in the trash bin, participants lose 10 points. Visual and auditive emphasize the changes in the player's score. We instruct players to score as much as possible. Errors of the AI can't be corrected by the AI. Once the player removes all spoiled and the robotic arm removed four pieces from the box, a screen in front of the players asks players to rate their experienced (see below) and confirm their selection by pressing and holding a virtual button in front of them. After that the next trial starts. \autoref{fig:Prototype} visualizes the task.

\subsection{Planned Procedure}
After providing written consent we ask participants to put on the VR headset HTC Vive Pro Eye as well as a Vive Tracker 2.0 device on the shoulder of their non-dominant hand. We then place participants seated into the center of the tracking area. After that the experiment starts. First we ask participants to run through a tutorial to get used to the control scheme of the game. We use one Valve Index Controller for the user input. Once the user feels comfortable with the game controls, the experiment starts and presenting the trials as prior described. 
\subsection{Conditions}
As mentioned earlier we will have two conditions in the experiment: One high performing and one low performing robot arm. The simulated AI of the robotic arm will not always grab correct. In some trials the robot arm will grab only 4 non-spoiled candies instead of the 4 spoiled ones. The condition of the round will determine how many of these error trials will happen: In the low-performance condition, the AI will have 70\% error trials during a complete round while only 20\% of the trials in the high-performance will be error trials. 

\subsection{Measurements}
We focus on the following aspects of in-game social behaviour to better understand the relationship between subjective trust measurement and the user's gaze behaviour:

\subsubsection{Rating and Questionnaire-based Measurement}
We use ratings to assess the situational trust of the participants after each trial as well as the participant's dispositional trust and trust in technology: 

\begin{itemize}
\item \textit{Dispositional Trust Level:} We used the  \citet{Wrightsman_1991} questionnaire in order to measure the participant's dispositional trust in others. It asks participants whether other individuals are honest and reliable and whether they consider them moral. The scale consists of 14 statements, seven positive and seven negative. Participants indicate their rating on a 7-point scale, ranging from -3 to 3, how much they agree with the shown statement. All answers are summed up to a score, ranging from -42 up to 42. This score indicates how much an individual may trust another person.\par
\item \textit{Trust in Technology:} Besides the assessment of trust in \textit{others}, we also assessed the user's level of trust in machines. We used the propensity to trust scale by \citet{Merritt2013ISystem}. This scale consists of six statements. Participants indicated their agreement with each statement on a five-point Likert scale, ranging from "strongly Agree" to "strongly Disagree".\par
\item \textit{Personal Self-esteem:} Since we were also curious about the role of general self-confidence, we used the \citet{Rosenberg1965RosenbergScale} self-esteem scale to assess the general self-confidence of the participants. The scale consists of 10 statements of both positive and negative feelings about oneself. Participants rated how much they agreed with these statements on a 4-point Likert scale, ranging from strongly disagree (1) to strongly agree (4). The sum of all items indicates self-esteem.\par
\item \textit{Experienced Trust of the AI}: After confirming the answer to the system, we asked participants to rate their experienced situational in the assistant using a Likert Scale from 0.0, not at all, to 1.0, very much in 100 steps.\par 
\item \textit{Demographic Information:} We asked the participants to indicate age and gender as well as to summarize the experiment to check for (lack of) attention paid to the study. \par
\end{itemize}

\subsubsection{Gaze Features}
As suggested by prior work \cite{Lu2019EyeReliability} we will analyze the following features:
\begin{itemize}
    \item \textbf{Fixation Characteristics}: We will measure the \textbf{total amount of time} (in seconds)  per trial where the user's gaze was focusing the robot arm as well as other parts in the game world.  We hypothesise that users who may experience low trust in the system will focus more the robot arm and less often other parts in the world.
    \item \textbf{Saccade Charactersitics}: We will investigate serveral aspects of the saccades: Building on prior work which suggests that low trust leads to a less organized and more spread gaze behaviour we want to analze several saccadic charactersitics, such as the mean saccade amplitude, the backtrack rate and the scanpath length. Furthermore we're curious about the spread of attention between the AI and the user's game field, which is why we plan to count the transition counts between the AI and the user's objects in the game. We hypothesis that low trust leads to more transitions between these two contexts. Furthermore we will analyze the velocity of saccadicintrusions, which are conjugate, horizontal micro-saccadic movements \cite{Biswas2018DetectingIntrusion}. As suggested by prior work, a increased cognitive workload increases also the average velocity of saccadicintrusions. Therefore we hypothesis that the increased amount of cognitive workload due to the experience of low trust may result in an increased velocity of these micro-saccadic movements.
    \item \textbf{Pupil Dilation and Blink Rate}: Additionally we plan to analyse the blink rate as well as the pupil dialation. As suggested by prior work a suppressed blink rate \cite{Paprocki2017WhatPerformance,Boehm-Davis2000TheWorkload} as well as pupil dilation \cite{Kret2015PupilPupils} may also be related to the user's cognitive workload. Building on these results we want to explore whether there is a connection between the experience of trust and the blinkrate as well as pupil dilation.
\end{itemize}

\section{Conclusion}
Trust is a core component for successful social interactions in our daily life with other humans, but also with digital solutions and tools. In recent years, researchers tried to look for ways to increase the user's trust in artificial intelligent systems. However, there are several limitations of currently discussed measurements, which mostly rely on using either standardized questionnaires, ratings or interviews after the user interacted with the system. Furthermore, these ways of measuring trust may not be applicable for measuring the user's trust during an interaction. On the other side, researchers suggest expensive and invasive sensors which may be used to continuously assess the user's trust in an interaction yet these sensors may feel too intrusive for users. However, measuring the user's gaze has been discussed as a way to bridge the gap between subjective measurements and sensory data. First evidence about the relationship between trust and attentional shifts were positive, other aspects of the user's gaze are yet untethered but may offer reliable ways to measure the user's trust. 
In this proposal, we suggest a study design to evaluate the relationship between the user's gaze and the subjective experience of trust and discuss which aspects of the user's gaze may reveal more insights about the experience of trust.

\bibliographystyle{ACM-Reference-Format}
\bibliography{references}


\begin{thebibliography}{52}


\ifx \showCODEN    \undefined \def \showCODEN     #1{\unskip}     \fi
\ifx \showDOI      \undefined \def \showDOI       #1{#1}\fi
\ifx \showISBNx    \undefined \def \showISBNx     #1{\unskip}     \fi
\ifx \showISBNxiii \undefined \def \showISBNxiii  #1{\unskip}     \fi
\ifx \showISSN     \undefined \def \showISSN      #1{\unskip}     \fi
\ifx \showLCCN     \undefined \def \showLCCN      #1{\unskip}     \fi
\ifx \shownote     \undefined \def \shownote      #1{#1}          \fi
\ifx \showarticletitle \undefined \def \showarticletitle #1{#1}   \fi
\ifx \showURL      \undefined \def \showURL       {\relax}        \fi
\providecommand\bibfield[2]{#2}
\providecommand\bibinfo[2]{#2}
\providecommand\natexlab[1]{#1}
\providecommand\showeprint[2][]{arXiv:#2}

\bibitem[Antony and Rowa(2005)]%
        {Antony2005}
\bibfield{author}{\bibinfo{person}{Martin~M. Antony} {and}
  \bibinfo{person}{Karen Rowa}.} \bibinfo{year}{2005}\natexlab{}.
\newblock \bibinfo{title}{{Evidence-based assessment of anxiety disorders in
  adults}}.
\newblock
\newblock
\showISSN{10403590}
\urldef\tempurl%
\url{https://doi.org/10.1037/1040-3590.17.3.256}
\showDOI{\tempurl}


\bibitem[Arrow(1974)]%
        {Arrow1974TheOrganization}
\bibfield{author}{\bibinfo{person}{Kenneth~J Arrow}.}
  \bibinfo{year}{1974}\natexlab{}.
\newblock \bibinfo{booktitle}{\emph{{The limits of organization}}}.
\newblock \bibinfo{publisher}{WW Norton {\&} Company}.
\newblock
\showISBNx{978-0393093230}


\bibitem[Barbato et~al\mbox{.}(2020)]%
        {Barbato2020TheOrienting}
\bibfield{author}{\bibinfo{person}{Mariapaola Barbato},
  \bibinfo{person}{Aisha~A. Almulla}, {and} \bibinfo{person}{Andrea Marotta}.}
  \bibinfo{year}{2020}\natexlab{}.
\newblock \showarticletitle{{The Effect of Trust on Gaze-Mediated Attentional
  Orienting}}.
\newblock \bibinfo{journal}{\emph{Frontiers in Psychology}}
  \bibinfo{volume}{11} (\bibinfo{date}{7} \bibinfo{year}{2020}).
\newblock
\showISSN{1664-1078}
\urldef\tempurl%
\url{https://doi.org/10.3389/fpsyg.2020.01554}
\showDOI{\tempurl}


\bibitem[Barreto et~al\mbox{.}(2008)]%
        {Barreto2008}
\bibfield{author}{\bibinfo{person}{Armando Barreto}, \bibinfo{person}{Ying
  Gao}, {and} \bibinfo{person}{Malek Adjouadi}.}
  \bibinfo{year}{2008}\natexlab{}.
\newblock \showarticletitle{{Pupil diameter measurements: untapped potential to
  enhance computer interaction for eye tracker users?}}
\newblock \bibinfo{journal}{\emph{Proceedings of the 10th international ACM
  SIGACCESS conference on Computers and accessibility}} (\bibinfo{year}{2008}),
  \bibinfo{pages}{269--270}.
\newblock
\showISBNx{978-1-59593-976-0}
\urldef\tempurl%
\url{https://doi.org/10.1145/1414471.1414532}
\showDOI{\tempurl}


\bibitem[Bawack et~al\mbox{.}(2022)]%
        {Bawack2022ArtificialReview}
\bibfield{author}{\bibinfo{person}{Ransome~Epie Bawack},
  \bibinfo{person}{Samuel~Fosso Wamba}, \bibinfo{person}{Kevin Daniel~André
  Carillo}, {and} \bibinfo{person}{Shahriar Akter}.}
  \bibinfo{year}{2022}\natexlab{}.
\newblock \showarticletitle{{Artificial intelligence in E-Commerce: a
  bibliometric study and literature review}}.
\newblock \bibinfo{journal}{\emph{Electronic Markets}} \bibinfo{volume}{32},
  \bibinfo{number}{1} (\bibinfo{date}{3} \bibinfo{year}{2022}),
  \bibinfo{pages}{297--338}.
\newblock
\showISSN{1019-6781}
\urldef\tempurl%
\url{https://doi.org/10.1007/s12525-022-00537-z}
\showDOI{\tempurl}


\bibitem[Bayliss and Tipper(2006)]%
        {Bayliss2006PredictiveYou}
\bibfield{author}{\bibinfo{person}{Andrew~P Bayliss} {and}
  \bibinfo{person}{Steven~P Tipper}.} \bibinfo{year}{2006}\natexlab{}.
\newblock \showarticletitle{{Predictive gaze cues and personality judgments:
  Should eye trust you?}}
\newblock \bibinfo{journal}{\emph{Psychological Science}} \bibinfo{volume}{17},
  \bibinfo{number}{6} (\bibinfo{year}{2006}), \bibinfo{pages}{514--520}.
\newblock
\showISSN{0956-7976}


\bibitem[Biswas and Prabhakar(2018)]%
        {Biswas2018DetectingIntrusion}
\bibfield{author}{\bibinfo{person}{Pradipta Biswas} {and}
  \bibinfo{person}{Gowdham Prabhakar}.} \bibinfo{year}{2018}\natexlab{}.
\newblock \showarticletitle{{Detecting drivers’ cognitive load from saccadic
  intrusion}}.
\newblock \bibinfo{journal}{\emph{Transportation Research Part F: Traffic
  Psychology and Behaviour}}  \bibinfo{volume}{54} (\bibinfo{year}{2018}),
  \bibinfo{pages}{63--78}.
\newblock
\showISSN{1369-8478}
\urldef\tempurl%
\url{https://doi.org/10.1016/j.trf.2018.01.017}
\showDOI{\tempurl}


\bibitem[B{\"{o}}ckle et~al\mbox{.}(2021)]%
        {Bockle2021CanInterfaces}
\bibfield{author}{\bibinfo{person}{Martin B{\"{o}}ckle}, \bibinfo{person}{Kwaku
  Yeboah-Antwi}, {and} \bibinfo{person}{Iana Kouris}.}
  \bibinfo{year}{2021}\natexlab{}.
\newblock \showarticletitle{{Can You Trust the Black Box? The Effect of
  Personality Traits on Trust in AI-Enabled User Interfaces}}. In
  \bibinfo{booktitle}{\emph{International Conference on Human-Computer
  Interaction}}. \bibinfo{publisher}{Springer}, \bibinfo{pages}{3--20}.
\newblock


\bibitem[Boehm-Davis et~al\mbox{.}(2000)]%
        {Boehm-Davis2000TheWorkload}
\bibfield{author}{\bibinfo{person}{Deborah~A Boehm-Davis},
  \bibinfo{person}{Wayne~D Gray}, {and} \bibinfo{person}{Michael~J Schoelles}.}
  \bibinfo{year}{2000}\natexlab{}.
\newblock \showarticletitle{{The Eye Blink as a Physiological Indicator of
  Cognitive Workload}}.
\newblock \bibinfo{journal}{\emph{Proceedings of the Human Factors and
  Ergonomics Society Annual Meeting}} \bibinfo{volume}{44},
  \bibinfo{number}{33} (\bibinfo{date}{7} \bibinfo{year}{2000}),
  \bibinfo{pages}{6--116}.
\newblock
\showISSN{2169-5067}
\urldef\tempurl%
\url{https://doi.org/10.1177/154193120004403309}
\showDOI{\tempurl}


\bibitem[Brzowski and Nathan-Roberts(2019)]%
        {Brzowski2019TrustReview}
\bibfield{author}{\bibinfo{person}{Matthew Brzowski} {and} \bibinfo{person}{Dan
  Nathan-Roberts}.} \bibinfo{year}{2019}\natexlab{}.
\newblock \showarticletitle{{Trust Measurement in Human–Automation
  Interaction: A Systematic Review}}.
\newblock \bibinfo{journal}{\emph{Proceedings of the Human Factors and
  Ergonomics Society Annual Meeting}} \bibinfo{volume}{63}, \bibinfo{number}{1}
  (\bibinfo{date}{11} \bibinfo{year}{2019}), \bibinfo{pages}{1595--1599}.
\newblock
\showISSN{2169-5067}
\urldef\tempurl%
\url{https://doi.org/10.1177/1071181319631462}
\showDOI{\tempurl}


\bibitem[Castelfranchi and Falcone(2010)]%
        {Castelfranchi2010_Socio}
\bibfield{author}{\bibinfo{person}{Christiano Castelfranchi} {and}
  \bibinfo{person}{Rino Falcone}.} \bibinfo{year}{2010}\natexlab{}.
\newblock \showarticletitle{{Socio-Cognitive Model of Trust: Basic
  Ingredients}}.
\newblock In \bibinfo{booktitle}{\emph{Trust Theory}}. \bibinfo{publisher}{John
  Wiley {\&} Sons, Ltd}, \bibinfo{address}{Chichester, UK}, Chapter~2,
  \bibinfo{pages}{35--94}.
\newblock
\urldef\tempurl%
\url{https://doi.org/10.1002/9780470519851.ch2}
\showDOI{\tempurl}


\bibitem[Chan-Olmsted(2019)]%
        {Chan-Olmsted2019AIndustry}
\bibfield{author}{\bibinfo{person}{Sylvia~M. Chan-Olmsted}.}
  \bibinfo{year}{2019}\natexlab{}.
\newblock \showarticletitle{{A Review of Artificial Intelligence Adoptions in
  the Media Industry}}.
\newblock \bibinfo{journal}{\emph{International Journal on Media Management}}
  \bibinfo{volume}{21}, \bibinfo{number}{3-4} (\bibinfo{date}{10}
  \bibinfo{year}{2019}), \bibinfo{pages}{193--215}.
\newblock
\showISSN{1424-1277}
\urldef\tempurl%
\url{https://doi.org/10.1080/14241277.2019.1695619}
\showDOI{\tempurl}


\bibitem[Cheng et~al\mbox{.}(2022)]%
        {Cheng2022HeartMetaanalysis}
\bibfield{author}{\bibinfo{person}{Ying‐Chih Cheng}, \bibinfo{person}{Min‐I
  Su}, \bibinfo{person}{Cheng‐Wei Liu}, \bibinfo{person}{Yu‐Chen Huang},
  {and} \bibinfo{person}{Wei‐Lieh Huang}.} \bibinfo{year}{2022}\natexlab{}.
\newblock \showarticletitle{{Heart rate variability in patients with anxiety
  disorders: A systematic review and meta‐analysis}}.
\newblock \bibinfo{journal}{\emph{Psychiatry and Clinical Neurosciences}}
  \bibinfo{volume}{76}, \bibinfo{number}{7} (\bibinfo{date}{7}
  \bibinfo{year}{2022}), \bibinfo{pages}{292--302}.
\newblock
\showISSN{1323-1316}
\urldef\tempurl%
\url{https://doi.org/10.1111/pcn.13356}
\showDOI{\tempurl}


\bibitem[Choung et~al\mbox{.}(2022)]%
        {Choung2022TrustTechnologiesb}
\bibfield{author}{\bibinfo{person}{Hyesun Choung}, \bibinfo{person}{Prabu
  David}, {and} \bibinfo{person}{Arun Ross}.} \bibinfo{year}{2022}\natexlab{}.
\newblock \showarticletitle{{Trust in AI and Its Role in the Acceptance of AI
  Technologies}}.
\newblock \bibinfo{journal}{\emph{International Journal of Human-Computer
  Interaction}} (\bibinfo{year}{2022}).
\newblock
\showISSN{15327590}
\urldef\tempurl%
\url{https://doi.org/10.1080/10447318.2022.2050543}
\showDOI{\tempurl}


\bibitem[Connor et~al\mbox{.}(2000)]%
        {Connor2000PsychometricScale}
\bibfield{author}{\bibinfo{person}{Kathryn~M Connor}, \bibinfo{person}{Jonathan
  R~T Davidson}, \bibinfo{person}{L~Erik Churchill}, \bibinfo{person}{Andrew
  Sherwood}, \bibinfo{person}{Richard~H Weisler}, {and} \bibinfo{person}{Edna
  Foa}.} \bibinfo{year}{2000}\natexlab{}.
\newblock \showarticletitle{{Psychometric properties of the Social Phobia
  Inventory (SPIN): New self-rating scale}}.
\newblock \bibinfo{journal}{\emph{British Journal of Psychiatry}}
  \bibinfo{volume}{176}, \bibinfo{number}{4} (\bibinfo{year}{2000}),
  \bibinfo{pages}{379--386}.
\newblock
\showISSN{0007-1250}
\urldef\tempurl%
\url{https://doi.org/DOI: 10.1192/bjp.176.4.379}
\showDOI{\tempurl}


\bibitem[Dorsey et~al\mbox{.}(2017)]%
        {Dorsey2017}
\bibfield{author}{\bibinfo{person}{E. Ray Dorsey}, \bibinfo{person}{Spyros
  Papapetropoulos}, \bibinfo{person}{Mulin Xiong}, {and} \bibinfo{person}{Karl
  Kieburtz}.} \bibinfo{year}{2017}\natexlab{}.
\newblock \showarticletitle{{The First Frontier: Digital Biomarkers for
  Neurodegenerative Disorders}}.
\newblock \bibinfo{journal}{\emph{Digital Biomarkers}} (\bibinfo{year}{2017}).
\newblock
\showISSN{2504-110X}
\urldef\tempurl%
\url{https://doi.org/10.1159/000477383}
\showDOI{\tempurl}


\bibitem[Du et~al\mbox{.}(2020)]%
        {Du2020PsychophysiologicalDriving}
\bibfield{author}{\bibinfo{person}{Na Du}, \bibinfo{person}{X.~Jessie Yang},
  {and} \bibinfo{person}{Feng Zhou}.} \bibinfo{year}{2020}\natexlab{}.
\newblock \showarticletitle{{Psychophysiological responses to takeover requests
  in conditionally automated driving}}.
\newblock \bibinfo{journal}{\emph{Accident Analysis {\&} Prevention}}
  \bibinfo{volume}{148} (\bibinfo{date}{12} \bibinfo{year}{2020}),
  \bibinfo{pages}{105804}.
\newblock
\showISSN{00014575}
\urldef\tempurl%
\url{https://doi.org/10.1016/j.aap.2020.105804}
\showDOI{\tempurl}


\bibitem[Eckstein et~al\mbox{.}(2017)]%
        {Eckstein2017BeyondDevelopment}
\bibfield{author}{\bibinfo{person}{Maria~K Eckstein}, \bibinfo{person}{Belén
  Guerra-Carrillo}, \bibinfo{person}{Alison~T Miller~Singley}, {and}
  \bibinfo{person}{Silvia~A Bunge}.} \bibinfo{year}{2017}\natexlab{}.
\newblock \showarticletitle{{Beyond eye gaze: What else can eyetracking reveal
  about cognition and cognitive development?}}
\newblock \bibinfo{journal}{\emph{Developmental Cognitive Neuroscience}}
  \bibinfo{volume}{25} (\bibinfo{year}{2017}), \bibinfo{pages}{69--91}.
\newblock
\showISSN{1878-9293}
\urldef\tempurl%
\url{https://doi.org/10.1016/j.dcn.2016.11.001}
\showDOI{\tempurl}


\bibitem[Ezer et~al\mbox{.}(2019)]%
        {Ezer2019TrustTeams}
\bibfield{author}{\bibinfo{person}{Neta Ezer}, \bibinfo{person}{Sylvain Bruni},
  \bibinfo{person}{Yang Cai}, \bibinfo{person}{Sam~J Hepenstal},
  \bibinfo{person}{Christopher~A Miller}, {and} \bibinfo{person}{Dylan~D
  Schmorrow}.} \bibinfo{year}{2019}\natexlab{}.
\newblock \showarticletitle{{Trust engineering for human-AI teams}}. In
  \bibinfo{booktitle}{\emph{Proceedings of the Human Factors and Ergonomics
  Society Annual Meeting}}, Vol.~\bibinfo{volume}{63}. \bibinfo{publisher}{SAGE
  Publications Sage CA: Los Angeles, CA}, \bibinfo{pages}{322--326}.
\newblock
\showISBNx{2169-5067}


\bibitem[Feng and Boyd-Graber(2019)]%
        {Feng2019WhatPlay}
\bibfield{author}{\bibinfo{person}{Shi Feng} {and} \bibinfo{person}{Jordan
  Boyd-Graber}.} \bibinfo{year}{2019}\natexlab{}.
\newblock \showarticletitle{{What can AI do for me?: evaluating machine
  learning interpretations in cooperative play}}. In
  \bibinfo{booktitle}{\emph{Proceedings of the 24th International Conference on
  Intelligent User Interfaces}}. \bibinfo{publisher}{ACM},
  \bibinfo{address}{New York, NY, USA}, \bibinfo{pages}{229--239}.
\newblock
\showISBNx{9781450362726}
\urldef\tempurl%
\url{https://doi.org/10.1145/3301275.3302265}
\showDOI{\tempurl}


\bibitem[Freeman(2018)]%
        {Freeman2018DoingHand}
\bibfield{author}{\bibinfo{person}{Jonathan~B. Freeman}.}
  \bibinfo{year}{2018}\natexlab{}.
\newblock \showarticletitle{{Doing Psychological Science by Hand}}.
\newblock \bibinfo{journal}{\emph{Current Directions in Psychological Science}}
  \bibinfo{volume}{27}, \bibinfo{number}{5} (\bibinfo{date}{10}
  \bibinfo{year}{2018}), \bibinfo{pages}{315--323}.
\newblock
\showISSN{0963-7214}
\urldef\tempurl%
\url{https://doi.org/10.1177/0963721417746793}
\showDOI{\tempurl}


\bibitem[Glikson and Woolley(2020)]%
        {Glikson2020HumanResearch}
\bibfield{author}{\bibinfo{person}{Ella Glikson} {and}
  \bibinfo{person}{Anita~Williams Woolley}.} \bibinfo{year}{2020}\natexlab{}.
\newblock \showarticletitle{{Human Trust in Artificial Intelligence: Review of
  Empirical Research}}.
\newblock \bibinfo{journal}{\emph{Academy of Management Annals}}
  \bibinfo{volume}{14}, \bibinfo{number}{2} (\bibinfo{date}{7}
  \bibinfo{year}{2020}), \bibinfo{pages}{627--660}.
\newblock
\showISSN{1941-6520}
\urldef\tempurl%
\url{https://doi.org/10.5465/annals.2018.0057}
\showDOI{\tempurl}


\bibitem[Grant and Spivey(2003)]%
        {Grant2003EyeThought}
\bibfield{author}{\bibinfo{person}{Elizabeth~R Grant} {and}
  \bibinfo{person}{Michael~J Spivey}.} \bibinfo{year}{2003}\natexlab{}.
\newblock \showarticletitle{{Eye movements and problem solving: Guiding
  attention guides thought}}.
\newblock \bibinfo{journal}{\emph{Psychological Science}} \bibinfo{volume}{14},
  \bibinfo{number}{5} (\bibinfo{year}{2003}), \bibinfo{pages}{462--466}.
\newblock
\showISSN{0956-7976}


\bibitem[Green et~al\mbox{.}(2007)]%
        {Green2007EyeAddition}
\bibfield{author}{\bibinfo{person}{Heather~J Green}, \bibinfo{person}{Patrick
  Lemaire}, {and} \bibinfo{person}{Stéphane Dufau}.}
  \bibinfo{year}{2007}\natexlab{}.
\newblock \showarticletitle{{Eye movement correlates of younger and older
  adults’ strategies for complex addition}}.
\newblock \bibinfo{journal}{\emph{Acta psychologica}} \bibinfo{volume}{125},
  \bibinfo{number}{3} (\bibinfo{year}{2007}), \bibinfo{pages}{257--278}.
\newblock
\showISSN{0001-6918}


\bibitem[Harrison~McKnight and Chervany(2001)]%
        {HarrisonMcKnight2001TrustTime}
\bibfield{author}{\bibinfo{person}{D. Harrison~McKnight} {and}
  \bibinfo{person}{Norman~L. Chervany}.} \bibinfo{year}{2001}\natexlab{}.
\newblock \showarticletitle{{Trust and Distrust Definitions: One Bite at a
  Time}}.
\newblock \bibinfo{pages}{27--54}.
\newblock
\urldef\tempurl%
\url{https://doi.org/10.1007/3-540-45547-7{\_}3}
\showDOI{\tempurl}


\bibitem[Johnson and Verdicchio(2017)]%
        {Johnson2017AIAnxiety}
\bibfield{author}{\bibinfo{person}{Deborah~G. Johnson} {and}
  \bibinfo{person}{Mario Verdicchio}.} \bibinfo{year}{2017}\natexlab{}.
\newblock \showarticletitle{{AI Anxiety}}.
\newblock \bibinfo{journal}{\emph{Journal of the Association for Information
  Science and Technology}} \bibinfo{volume}{68}, \bibinfo{number}{9}
  (\bibinfo{date}{9} \bibinfo{year}{2017}), \bibinfo{pages}{2267--2270}.
\newblock
\showISSN{23301635}
\urldef\tempurl%
\url{https://doi.org/10.1002/asi.23867}
\showDOI{\tempurl}


\bibitem[Koffskey et~al\mbox{.}(2014)]%
        {Koffskey2014UsingOperators}
\bibfield{author}{\bibinfo{person}{Christina Koffskey},
  \bibinfo{person}{Laura~H. Ikuma}, \bibinfo{person}{Craig Harvey}, {and}
  \bibinfo{person}{Fereydoun Aghazadeh}.} \bibinfo{year}{2014}\natexlab{}.
\newblock \showarticletitle{{Using eye-tracking to investigate strategy and
  performance of expert and novice control room operators}}. In
  \bibinfo{booktitle}{\emph{Proceedings of the Human Factors and Ergonomics
  Society}}.
\newblock
\showISBNx{9780945289456}
\showISSN{10711813}
\urldef\tempurl%
\url{https://doi.org/10.1177/1541931214581348}
\showDOI{\tempurl}


\bibitem[Kret et~al\mbox{.}(2015)]%
        {Kret2015PupilPupils}
\bibfield{author}{\bibinfo{person}{M~E Kret}, \bibinfo{person}{A~H Fischer},
  {and} \bibinfo{person}{C~K~W De~Dreu}.} \bibinfo{year}{2015}\natexlab{}.
\newblock \showarticletitle{{Pupil Mimicry Correlates With Trust in In-Group
  Partners With Dilating Pupils}}.
\newblock \bibinfo{journal}{\emph{Psychological Science}} \bibinfo{volume}{26},
  \bibinfo{number}{9} (\bibinfo{date}{7} \bibinfo{year}{2015}),
  \bibinfo{pages}{1401--1410}.
\newblock
\showISSN{0956-7976}
\urldef\tempurl%
\url{https://doi.org/10.1177/0956797615588306}
\showDOI{\tempurl}


\bibitem[Lachman and Joffe(2021)]%
        {Lachman2021ApplicationsEntertainment}
\bibfield{author}{\bibinfo{person}{Richard Lachman} {and}
  \bibinfo{person}{Michael Joffe}.} \bibinfo{year}{2021}\natexlab{}.
\newblock \showarticletitle{{Applications of Artificial Intelligence in Media
  and Entertainment}}.
\newblock \bibinfo{pages}{201--220}.
\newblock
\urldef\tempurl%
\url{https://doi.org/10.4018/978-1-7998-3499-1.ch012}
\showDOI{\tempurl}


\bibitem[Lai et~al\mbox{.}(2013)]%
        {Lai2013A2012}
\bibfield{author}{\bibinfo{person}{Meng-Lung Lai}, \bibinfo{person}{Meng-Jung
  Tsai}, \bibinfo{person}{Fang-Ying Yang}, \bibinfo{person}{Chung-Yuan Hsu},
  \bibinfo{person}{Tzu-Chien Liu}, \bibinfo{person}{Silvia Wen-Yu Lee},
  \bibinfo{person}{Min-Hsien Lee}, \bibinfo{person}{Guo-Li Chiou},
  \bibinfo{person}{Jyh-Chong Liang}, {and} \bibinfo{person}{Chin-Chung Tsai}.}
  \bibinfo{year}{2013}\natexlab{}.
\newblock \showarticletitle{{A review of using eye-tracking technology in
  exploring learning from 2000 to 2012}}.
\newblock \bibinfo{journal}{\emph{Educational research review}}
  \bibinfo{volume}{10} (\bibinfo{year}{2013}), \bibinfo{pages}{90--115}.
\newblock
\showISSN{1747-938X}


\bibitem[Liu(2010)]%
        {Liu2010Human-MachineInteraction}
\bibfield{author}{\bibinfo{person}{Conghui Liu}.}
  \bibinfo{year}{2010}\natexlab{}.
\newblock \showarticletitle{{Human-Machine Trust Interaction}}.
\newblock \bibinfo{journal}{\emph{International Journal of Dependable and
  Trustworthy Information Systems}} \bibinfo{volume}{1}, \bibinfo{number}{4}
  (\bibinfo{date}{10} \bibinfo{year}{2010}), \bibinfo{pages}{61--74}.
\newblock
\showISSN{1947-9050}
\urldef\tempurl%
\url{https://doi.org/10.4018/jdtis.2010100104}
\showDOI{\tempurl}


\bibitem[Loo(2002)]%
        {Loo2002AScales}
\bibfield{author}{\bibinfo{person}{Robert Loo}.}
  \bibinfo{year}{2002}\natexlab{}.
\newblock \showarticletitle{{A caveat on using single‐item versus
  multiple‐item scales}}.
\newblock \bibinfo{journal}{\emph{Journal of Managerial Psychology}}
  \bibinfo{volume}{17}, \bibinfo{number}{1} (\bibinfo{date}{2}
  \bibinfo{year}{2002}), \bibinfo{pages}{68--75}.
\newblock
\showISSN{0268-3946}
\urldef\tempurl%
\url{https://doi.org/10.1108/02683940210415933}
\showDOI{\tempurl}


\bibitem[Lu and Sarter(2019)]%
        {Lu2019EyeReliability}
\bibfield{author}{\bibinfo{person}{Y Lu} {and} \bibinfo{person}{N Sarter}.}
  \bibinfo{year}{2019}\natexlab{}.
\newblock \showarticletitle{{Eye Tracking: A Process-Oriented Method for
  Inferring Trust in Automation as a Function of Priming and System
  Reliability}}.
\newblock \bibinfo{journal}{\emph{IEEE Transactions on Human-Machine Systems}}
  \bibinfo{volume}{49}, \bibinfo{number}{6} (\bibinfo{year}{2019}),
  \bibinfo{pages}{560--568}.
\newblock
\showISSN{2168-2305}
\urldef\tempurl%
\url{https://doi.org/10.1109/THMS.2019.2930980}
\showDOI{\tempurl}


\bibitem[Luna and Velanova(2011)]%
        {Luna2011DevelopmentMovements}
\bibfield{author}{\bibinfo{person}{Beatriz Luna} {and}
  \bibinfo{person}{Katerina Velanova}.} \bibinfo{year}{2011}\natexlab{}.
\newblock \showarticletitle{{Development from reflexive to controlled eye
  movements}}.
\newblock  (\bibinfo{year}{2011}).
\newblock


\bibitem[Merritt et~al\mbox{.}(2013)]%
        {Merritt2013ISystem}
\bibfield{author}{\bibinfo{person}{Stephanie~M. Merritt},
  \bibinfo{person}{Heather Heimbaugh}, \bibinfo{person}{Jennifer LaChapell},
  {and} \bibinfo{person}{Deborah Lee}.} \bibinfo{year}{2013}\natexlab{}.
\newblock \showarticletitle{{I Trust It, but I Don’t Know Why: Effects of
  Implicit Attitudes Toward Automation on Trust in an Automated System}}.
\newblock \bibinfo{journal}{\emph{Human Factors: The Journal of the Human
  Factors and Ergonomics Society}} \bibinfo{volume}{55}, \bibinfo{number}{3}
  (\bibinfo{date}{6} \bibinfo{year}{2013}), \bibinfo{pages}{520--534}.
\newblock
\showISSN{0018-7208}
\urldef\tempurl%
\url{https://doi.org/10.1177/0018720812465081}
\showDOI{\tempurl}


\bibitem[Nazareth and Kim(2020)]%
        {Nazareth2020TheTechnology}
\bibfield{author}{\bibinfo{person}{Roland~Paul Nazareth} {and}
  \bibinfo{person}{Jung~Hyup Kim}.} \bibinfo{year}{2020}\natexlab{}.
\newblock \showarticletitle{{The Impact of Eye Tracking Technology}}. In
  \bibinfo{booktitle}{\emph{Advances in Usability, User Experience, Wearable
  and Assistive Technology}}, \bibfield{editor}{\bibinfo{person}{Tareq Ahram}
  {and} \bibinfo{person}{Christianne Falc{\~{a}}o}} (Eds.).
  \bibinfo{publisher}{Springer International Publishing},
  \bibinfo{address}{Cham}, \bibinfo{pages}{524--530}.
\newblock
\showISBNx{978-3-030-51828-8}


\bibitem[Papenmeier et~al\mbox{.}(2022)]%
        {Papenmeier2022ItsAI}
\bibfield{author}{\bibinfo{person}{Andrea Papenmeier}, \bibinfo{person}{Dagmar
  Kern}, \bibinfo{person}{Gwenn Englebienne}, {and} \bibinfo{person}{Christin
  Seifert}.} \bibinfo{year}{2022}\natexlab{}.
\newblock \showarticletitle{{It’s Complicated: The Relationship between User
  Trust, Model Accuracy and Explanations in AI}}.
\newblock \bibinfo{journal}{\emph{ACM Transactions on Computer-Human
  Interaction}} \bibinfo{volume}{29}, \bibinfo{number}{4} (\bibinfo{date}{8}
  \bibinfo{year}{2022}), \bibinfo{pages}{1--33}.
\newblock
\showISSN{1073-0516}
\urldef\tempurl%
\url{https://doi.org/10.1145/3495013}
\showDOI{\tempurl}


\bibitem[Paprocki and Lenskiy(2017)]%
        {Paprocki2017WhatPerformance}
\bibfield{author}{\bibinfo{person}{Rafal Paprocki} {and} \bibinfo{person}{Artem
  Lenskiy}.} \bibinfo{year}{2017}\natexlab{}.
\newblock \showarticletitle{{What Does Eye-Blink Rate Variability Dynamics Tell
  Us About Cognitive Performance?}}
\newblock \bibinfo{journal}{\emph{Frontiers in Human Neuroscience}}
  \bibinfo{volume}{11} (\bibinfo{date}{12} \bibinfo{year}{2017}).
\newblock
\showISSN{1662-5161}
\urldef\tempurl%
\url{https://doi.org/10.3389/fnhum.2017.00620}
\showDOI{\tempurl}


\bibitem[Rayner(1998)]%
        {Rayner1998EyeResearch.}
\bibfield{author}{\bibinfo{person}{Keith Rayner}.}
  \bibinfo{year}{1998}\natexlab{}.
\newblock \showarticletitle{{Eye movements in reading and information
  processing: 20 years of research.}}
\newblock \bibinfo{journal}{\emph{Psychological bulletin}}
  \bibinfo{volume}{124}, \bibinfo{number}{3} (\bibinfo{year}{1998}),
  \bibinfo{pages}{372}.
\newblock
\showISSN{1939-1455}


\bibitem[Rosenberg(1965)]%
        {Rosenberg1965RosenbergScale}
\bibfield{author}{\bibinfo{person}{Morris Rosenberg}.}
  \bibinfo{year}{1965}\natexlab{}.
\newblock \showarticletitle{{Rosenberg self-esteem scale}}.
\newblock \bibinfo{journal}{\emph{Journal of Religion and Health}}
  (\bibinfo{year}{1965}).
\newblock


\bibitem[Shehab and Abdulle(2011)]%
        {Shehab2011CognitiveHypertension}
\bibfield{author}{\bibinfo{person}{Abdullah Shehab} {and}
  \bibinfo{person}{Abdishakur Abdulle}.} \bibinfo{year}{2011}\natexlab{}.
\newblock \showarticletitle{{Cognitive and autonomic dysfunction measures in
  normal controls, white coat and borderline hypertension}}.
\newblock \bibinfo{journal}{\emph{BMC Cardiovascular Disorders}}
  (\bibinfo{year}{2011}).
\newblock
\showISSN{14712261}
\urldef\tempurl%
\url{https://doi.org/10.1186/1471-2261-11-3}
\showDOI{\tempurl}


\bibitem[Spence and Rapee(2016)]%
        {Spence2016}
\bibfield{author}{\bibinfo{person}{Susan~H. Spence} {and}
  \bibinfo{person}{Ronald~M. Rapee}.} \bibinfo{year}{2016}\natexlab{}.
\newblock \bibinfo{title}{{The etiology of social anxiety disorder: An
  evidence-based model}}.
\newblock
\newblock
\showISSN{1873622X}
\urldef\tempurl%
\url{https://doi.org/10.1016/j.brat.2016.06.007}
\showDOI{\tempurl}


\bibitem[Strimbu and Tavel(2010)]%
        {Strimbu2010}
\bibfield{author}{\bibinfo{person}{Kyle Strimbu} {and} \bibinfo{person}{Jorge~A
  Tavel}.} \bibinfo{year}{2010}\natexlab{}.
\newblock \showarticletitle{{What are biomarkers?}}
\newblock \bibinfo{journal}{\emph{Current opinion in HIV and AIDS}}
  \bibinfo{volume}{5}, \bibinfo{number}{6} (\bibinfo{date}{11}
  \bibinfo{year}{2010}), \bibinfo{pages}{463--466}.
\newblock
\showISSN{1746-6318}
\urldef\tempurl%
\url{https://doi.org/10.1097/COH.0b013e32833ed177}
\showDOI{\tempurl}


\bibitem[Theeuwes et~al\mbox{.}(2009)]%
        {Theeuwes2009InteractionsMovements}
\bibfield{author}{\bibinfo{person}{Jan Theeuwes}, \bibinfo{person}{Artem
  Belopolsky}, {and} \bibinfo{person}{Christian N~L Olivers}.}
  \bibinfo{year}{2009}\natexlab{}.
\newblock \showarticletitle{{Interactions between working memory, attention and
  eye movements}}.
\newblock \bibinfo{journal}{\emph{Acta psychologica}} \bibinfo{volume}{132},
  \bibinfo{number}{2} (\bibinfo{year}{2009}), \bibinfo{pages}{106--114}.
\newblock
\showISSN{0001-6918}


\bibitem[Vereschak et~al\mbox{.}(2021)]%
        {Vereschak2021HowMethodologies}
\bibfield{author}{\bibinfo{person}{Oleksandra Vereschak},
  \bibinfo{person}{Gilles Bailly}, {and} \bibinfo{person}{Baptiste Caramiaux}.}
  \bibinfo{year}{2021}\natexlab{}.
\newblock \showarticletitle{{How to Evaluate Trust in AI-Assisted Decision
  Making? A Survey of Empirical Methodologies}}.
\newblock \bibinfo{journal}{\emph{Proceedings of the ACM on Human-Computer
  Interaction}} \bibinfo{volume}{5}, \bibinfo{number}{CSCW2}
  (\bibinfo{date}{10} \bibinfo{year}{2021}).
\newblock
\showISSN{25730142}
\urldef\tempurl%
\url{https://doi.org/10.1145/3476068}
\showDOI{\tempurl}


\bibitem[Wrightsman(1991)]%
        {Wrightsman_1991}
\bibfield{author}{\bibinfo{person}{Lawrence~S Wrightsman}.}
  \bibinfo{year}{1991}\natexlab{}.
\newblock \showarticletitle{{Interpersonal trust and attitudes toward human
  nature.}}
\newblock  (\bibinfo{year}{1991}).
\newblock


\bibitem[Yarbus and Yarbus(1967)]%
        {Yarbus1967EyeObjects}
\bibfield{author}{\bibinfo{person}{Alfred~L Yarbus} {and}
  \bibinfo{person}{Alfred~L Yarbus}.} \bibinfo{year}{1967}\natexlab{}.
\newblock \showarticletitle{{Eye movements during perception of complex
  objects}}.
\newblock \bibinfo{journal}{\emph{Eye movements and vision}}
  (\bibinfo{year}{1967}), \bibinfo{pages}{171--211}.
\newblock
\showISSN{1489953817}


\bibitem[Yi et~al\mbox{.}(2020)]%
        {Yi2020IdentificationEmotions}
\bibfield{author}{\bibinfo{person}{Qian Yi}, \bibinfo{person}{Shiquan Xiong},
  \bibinfo{person}{Biao Wang}, {and} \bibinfo{person}{Shuping Yi}.}
  \bibinfo{year}{2020}\natexlab{}.
\newblock \showarticletitle{{Identification of trusted interactive behavior
  based on mouse behavior considering web User's emotions}}.
\newblock \bibinfo{journal}{\emph{International Journal of Industrial
  Ergonomics}}  \bibinfo{volume}{76} (\bibinfo{date}{3} \bibinfo{year}{2020}),
  \bibinfo{pages}{102903}.
\newblock
\showISSN{01698141}
\urldef\tempurl%
\url{https://doi.org/10.1016/j.ergon.2019.102903}
\showDOI{\tempurl}


\bibitem[Yin et~al\mbox{.}(2019)]%
        {Yin2019UnderstandingModels}
\bibfield{author}{\bibinfo{person}{Ming Yin}, \bibinfo{person}{Jennifer
  Wortman~Vaughan}, {and} \bibinfo{person}{Hanna Wallach}.}
  \bibinfo{year}{2019}\natexlab{}.
\newblock \showarticletitle{{Understanding the Effect of Accuracy on Trust in
  Machine Learning Models}}. In \bibinfo{booktitle}{\emph{Proceedings of the
  2019 CHI Conference on Human Factors in Computing Systems}}.
  \bibinfo{publisher}{ACM}, \bibinfo{address}{New York, NY, USA},
  \bibinfo{pages}{1--12}.
\newblock
\showISBNx{9781450359702}
\urldef\tempurl%
\url{https://doi.org/10.1145/3290605.3300509}
\showDOI{\tempurl}


\bibitem[Yu et~al\mbox{.}(2018)]%
        {Yu2018ArtificialHealthcare}
\bibfield{author}{\bibinfo{person}{Kun-Hsing Yu}, \bibinfo{person}{Andrew~L.
  Beam}, {and} \bibinfo{person}{Isaac~S. Kohane}.}
  \bibinfo{year}{2018}\natexlab{}.
\newblock \showarticletitle{{Artificial intelligence in healthcare}}.
\newblock \bibinfo{journal}{\emph{Nature Biomedical Engineering}}
  \bibinfo{volume}{2}, \bibinfo{number}{10} (\bibinfo{date}{10}
  \bibinfo{year}{2018}), \bibinfo{pages}{719--731}.
\newblock
\showISSN{2157-846X}
\urldef\tempurl%
\url{https://doi.org/10.1038/s41551-018-0305-z}
\showDOI{\tempurl}


\bibitem[Yuksel et~al\mbox{.}(2017)]%
        {Yuksel2017BrainsInteractions}
\bibfield{author}{\bibinfo{person}{Beste~F. Yuksel}, \bibinfo{person}{Penny
  Collisson}, {and} \bibinfo{person}{Mary Czerwinski}.}
  \bibinfo{year}{2017}\natexlab{}.
\newblock \showarticletitle{{Brains or Beauty: How to Engender Trust in
  User-Agent Interactions}}.
\newblock \bibinfo{journal}{\emph{ACM Transactions on Internet Technology}}
  \bibinfo{volume}{17}, \bibinfo{number}{1} (\bibinfo{date}{3}
  \bibinfo{year}{2017}), \bibinfo{pages}{1--20}.
\newblock
\showISSN{1533-5399}
\urldef\tempurl%
\url{https://doi.org/10.1145/2998572}
\showDOI{\tempurl}


\bibitem[Zhang et~al\mbox{.}(2020)]%
        {Zhang2020HumanReview}
\bibfield{author}{\bibinfo{person}{Ruohan Zhang}, \bibinfo{person}{Akanksha
  Saran}, \bibinfo{person}{Bo Liu}, \bibinfo{person}{Yifeng Zhu},
  \bibinfo{person}{Sihang Guo}, \bibinfo{person}{Scott Niekum},
  \bibinfo{person}{Dana Ballard}, {and} \bibinfo{person}{Mary Hayhoe}.}
  \bibinfo{year}{2020}\natexlab{}.
\newblock \showarticletitle{{Human Gaze Assisted Artificial Intelligence: A
  Review}}.
\newblock \bibinfo{journal}{\emph{IJCAI : proceedings of the conference}}
  \bibinfo{volume}{2020} (\bibinfo{date}{7} \bibinfo{year}{2020}),
  \bibinfo{pages}{4951--4958}.
\newblock
\showISSN{1045-0823}
\urldef\tempurl%
\url{https://doi.org/10.24963/ijcai.2020/689}
\showDOI{\tempurl}


\end{thebibliography}

\end{document}